\documentstyle[preprint,prl,aps]{revtex} 
\begin{document}
\draft
\title{ Multi-anyons in the magnetic field}
\author{Taeseung Choi$^{a}$, Chang-Mo Ryu$^{a}$ and Chul Koo Kim$^{b}$}
\address{$^{a}$ Department of Physics, Pohang University of Science 
and Technology, Pohang 790-784, Korea \\
$^{b}$ Department of Physics, Yonsei University,
         Seoul 120-749, Korea}
\date{\today}

\maketitle

\begin{abstract}
We consider the external magnetic field effects on the two types of
anyon with fractional statistical parameters $p/q$ with
coprimes $p$ and $q$, one with fractional charge $e/q $ and
flux $p \phi_0(=hc/e)$(type I), the other with fractional flux
$p \phi_0/q$ and fundamental charge $e$(type II).
These two-types of anyons show different behaviors in the presence of 
the external magnetic field.
We also considered the geometry in which a two-dimensional plane 
contains an island of anyons with different statistical parameter
in their equilibrium.
The equilibrium inside an island is shown to be periodic with respect
to the flux through the island. 
The period for the type I anyon equals to the
integer multiple of the fundamental flux quantum.
In the case of type II anyon the period is found to be 
the fractional multiple of the fundamental flux quantum.

\end{abstract}

In two spatial dimension it is possible to interpolate the statistics 
between the bosonic and fermionic cases \cite{Lein,Wilc1}. 
F. Wilczek called the particle with this exotic statistics, 
"anyon" \cite{Wilc2}. 
Anyons play a crucial role in understanding the certain 
two-dimensional electron systems, most notably the fractional quantum 
Hall effect \cite{Laugh,Girv1}. 
Anyons can be considered as bosons carrying a fictitious charges 
$Q$ and a fictitious flux tube with impenetrable flux $\phi$, 
mutually interacting via Aharonov-Bohm-type couplings, 
and such that the fictitious charge and flux 
are related to the statistical angle $\theta$ ($=  
\phi / \phi^*_0 $, where $\phi^*_0 = hc/Q$). 
The exchange of two anyons gives the phase $\exp\left[i \theta \pi\right]$.
We will consider the two types of anyons with the same statistical
parameter $\theta (=p/q, $ $ p$ and $q$ are coprimes), one with the 
charge $e^* (= e/q )$ and the fictitious flux $p \phi_0$, the other with
the charge $e$ and the fractional flux $p \phi_0 /q$. 
Where $\phi_0$ is the fundamental flux $hc/e$.
We will call the former as type I and the latter as type II in the following.
The type I anyon is believed to be the elementary excitation in the 
$\nu$ ($=p^{'} /q$) fractional quantum Hall liquid \cite{Laugh,Halp}, 
where $pp^{'}=1$ mod $q$ \cite{Moor}.
In the previous work \cite{Ours} we considered the persistent currents
for the type II anyons of the same statistical angle with the charge $e$
and fractional flux $p \phi_0 /q$.
This is the quasiparticle around the chiral spin liquids \cite{Wen}.
These two types are indistinguishable when there is no interaction 
with external electromagnetic (EM) fields. 
However, we show that, since the external EM-field is minimally 
coupled to charged particles with a coupling constant proportional to 
their charge, 
the two types display different behaviors under the external EM-field.

We first consider the Hamiltonian of $N$ anyons of type I (charge 
$e^*(>0)$, flux $p \phi_0$) in an external magnetic field given by
\begin{equation}
H= \frac{1}{2m} \sum_{n=1}^N \bigg[ \bigg({\bf p}_n - 
\frac{e^*}{c} {\bf A}_n( {\bf r}_1,\cdots,{\bf r}_N ) 
\bigg)^2 -\frac{e^* \hbar}{2mc} \sigma_n 
{\bf B}_n( {\bf r}_1,\cdots,{\bf r}_N ) \bigg],
\end{equation}
where $\sigma_n(\pm)$ represents the spin of $n$th particle and
${\bf B}_n( {\bf r}_1,\cdots,{\bf r}_N )$ is the magnetic field
seen by the $n$th particle. 
The gauge potential ${\bf A}_n$ corresponding to ${\bf B}_n$
is assumed to be of the form
$ {\bf A}_n( {\bf r}_1,\cdots,{\bf r}_N )
= {\bf A}_{ex}( {\bf r}_n ) +
\sum_{m (\ne n)} {\bf A}_{nm}( {\bf r}_n - {\bf r}_m )$.
The gauge potential ${\bf A}_{nm} ( {\bf r}_n - {\bf r}_m )$
 which gives the statistical interaction 
 $$
 A_{nm}^i ( {\bf r}_n - {\bf r}_m )=
- \theta \frac{\epsilon_{ij} (r_n^j - r_m^j)}
 {|  {\bf r}_n - {\bf r}_m |^2}.
 $$
 And the external gauge potential ${\bf A}_{ex}$ is
 composed of ${\bf A}_c + {\bf A}_s$.
 Where the ${\bf A}_c$ corresponds to a uniform magnetic field 
 $B (>0)$ and the ${\bf A}_s$ correspons to the singular
 magnetic flux $\phi_s$ at the origin. They are defined as
 $$
 A_s^i ({\bf r}_n) = - \phi_s \frac{\epsilon_{ij} r_n^j}
 {2 \pi |r_n|^2} , ~~~\mbox{and} ~~~~
 A_c^i ({\bf r}_n) = -\frac{\epsilon_{ij}}{4} 
 \frac{\partial}{\partial r_n^j} B |r_n|^2
 $$
 The singular flux at the origin gives rise to the AB geometry.
The general arguments about the exact $N$-body ground state of 
anyonlike objects was discussed in Ref. \cite{Choi}.
There, anyons have been considered as fermions with generic two-body 
flux-tube interactions. In Ref. \cite{Choi} they considered the 
anyonlike particles of charge $-e$ with arbitrary flux. 
In our case we consider the anyons of fractional charge with the 
integer multiple of fundamental flux. 
So the resulting wavefunction becomes slightly different from 
the ones in Ref. \cite{Ours}. 
We shall briefly review some known results of the exact $N$-body ground
state for our purpose.
Using the singular gauge transformation the multivalued wavefunction
$\Psi_s$ is related to the original wave function $\Psi_o$ by 
the transformation
\begin{eqnarray}
\Psi_s({\bf r}_1, \cdots, {\bf r}_N) &=& 
\Omega \Psi_o({\bf r}_1, \cdots, {\bf r}_N), \\ \nonumber
\Omega&=& \prod_n e^{ i \alpha_s \Theta_n} 
\prod_{n<m} e^{i \theta \Theta_{nm}}, \\ \nonumber
\tan{\Theta_n} & \equiv & \frac{y_n}{x_n}, ~~~~\mbox{and}~~~~~
\tan{\Theta_{nm}} \equiv \frac{y_n -y_m}{x_n -x_m},
\end{eqnarray}
where $\alpha_s =  \phi_s/\phi_0^*(= \phi_s / (q \phi_0))$.  
Since the flux is impenetrable, all the particles cannot 
be in the same position and moreover cannot be located at the origin.
That is to say, the following boundary condition is satisfied
\begin{equation}
\lim_{{\bf r}_i \rightarrow {\bf 0}} 
\Psi_s ({\bf r}_1 \cdots, {\bf r}_N) =0, ~~ 
\lim_{{\bf r}_i \rightarrow {\bf r}_j} 
\Psi_s ({\bf r}_1 \cdots, {\bf r}_N) =0,
\end{equation}
for arbitrary $1<i,j<N$ and $i \ne j$.
Hence only the uniform magnetic field contributes to the Zeeman term in the 
Hamiltonian.  And $\Psi_s$ satisfies the eigenequation due to $H_S$ 
(${\bf k}$ is the unit vector perpendicular to the plane 
${\bf r}_{ij} = {\bf r}_i - {\bf r}_j$).
\begin{equation}
H_s=\Omega H \Omega^{-1}=
 \frac{1}{2m} \sum_{n=1}^{N} \bigg[ \bigg( {\bf p}_n - 
\frac{e^* B}{2} {\bf k} \times {\bf r}_n  \bigg)^2 
- \frac{e^* \hbar}{2 mc} \sigma_n B \bigg].
\end{equation}
We note that the total (canonical) angular momentum operator 
$J_t ( = -\sum_n i \hbar \partial_{\Theta_n})$
commutes with $H_s$ so that the total angular momentum of this
system is a good quantum number.

This Hamiltonian can be factored into \cite{Choi,Girv2}
\begin{equation}
H_s= - \sum_{n=1}^N \frac{\hbar^2}{2 m}
(D_n^x - i \sigma_n D_n^y)(D_n^x +i \sigma_n D_n^y),
\end{equation}
where $D_n^{x,y} \equiv \partial_n^{x,y} - i \frac{e^*}{\hbar c}
{A_c}^{x,y} ({\bf r}_n)$,
and  ${A_c}^{x,y} ({\bf r}_n)$.
Then the ground state is defined by
\begin{equation}
(D_n^x + i \sigma_n D_n^y) \Psi_s ({\bf r}_1,\cdots,{\bf r}_N )=0, ~~
n=1, \cdots, N,
\end{equation}
since $H_s$ is non-negative definite.
With the substitution
\begin{equation}
\Psi_s({\bf r}_1,\cdots,{\bf r}_N)=f({\bf r}_1,\cdots,{\bf r}_N)
\exp\left[-\frac{\sigma_n \pi B}{2 \phi^*_0}
\sum_n |{\bf r}_n|^2\right].
\end{equation}
Equation (6) becomes 
$$
\bigg[ \frac{\partial}{\partial x_n} + i \sigma_n 
\frac{\partial}{\partial y_n} \bigg] f({\bf r}_1, \cdots, {\bf r}_N)
=0, ~~~~ n=1, \cdots, N,
$$
implying that $f({\bf r}_1,\cdots,{\bf r}_N)$ is an entire function of 
$z_n \equiv x_n + i \sigma_n y_n$ except the points 
${\bf r}_n={\bf r}_m$ for all $n \ne m$ and ${\bf r}_n={\bf 0}$ for 
all $n$. 
These points are excluded by the boundary condition (4).
For the wavefunction to be physically meaningful, we must take into
account the normalizabily of $\Psi_s$.  In Eq. (7) the normalization 
of $\Psi_s$ is possible only for $\sigma_n =+1$, which we will assume 
in the remaining part of this paper.
Accounting for the multivaluedness, the most general form of
the ground-state wave function becomes the following entire functions, 
\begin{equation}
\Psi_s (z_1, \cdots, z_N)= \prod_i z_i^{ \alpha_s} 
\prod_{i < j} (z_i - z_j)^{ \theta} {\cal S}\prod_i z_i^{ n^{'}_i}  
\prod_{i < j} (z_i - z_j)^{ k_{ij}}   
\exp\left[- \frac{2 \pi B}{4 \phi^*_0} \sum_i |z_i|^2 \right],
\end{equation}
only in the region where the boundary condition is satisfied,
where ${\cal S}$ is the symmetrizer and $k_{ij}$ is an even integer.
The charge of the type I anyon is $e/q$ so that the simple minded
gauge invariance gives the multiple period $q \phi_0$.
But it seems to be incompatible with the very
general argument of Byers and Yang \cite{Byers}. The real constituents
are electrons, so the period must always be the fundamental period 
$\phi_0=hc/e$, regardless of the nature of the electron-electron 
interaction. The simple and reasonable interpretation is that the 
original wave function must be multivalued in any true AB geometry, 
following Kivelson and Rocek \cite{Kive1}.
The integer $n^{'}_i$ becomes fractional number $n_i/q$ for the
multivalued wave function, since it returns to the original state
for $q$ turns around the impenetrable flux tube even for $\phi_s=0$.
The total angular momentum of the wave function
is $N \alpha_s + \theta N(N-1)/2 +\sum_i n^{'}_i +
\sum_{i<j} k_{ij}$.  
This total angular momentum returns to the same value for the shift
$n^{'}_i \rightarrow n^{'}_i - 1/q$, when the singular AB
flux at the origin increased by one flux quantum $\phi_0$, i.e.,
$\alpha_c \rightarrow \alpha_c + 1/q$.
Since the eigenvalues of both the energy and the total angular momentum
is periodic with the period of one flux quantum,
the system displays the fundamental periodicity with respect to the
AB flux, as we expected by the gauge invariance argument.

The terms on the left of ${\cal S}$ represents the multivaluedness of 
$\Psi_s$ which comes from the singular gauge transformation Eq. (2).
The boundary condition (3) gives the constraint to $\Psi_s$ such
that the exponent of $z_i$ and $z_i -z_j$ must be positive.
The normalizability of the wave function is automatically 
satisfied by the 
$\exp \left[- \frac{2 \pi B}{4 \phi^*_0} \sum_i |z_i|^2\right]$ 
term when the magnetic field exists at the infinity.
But in the case the uniform magnetic field is confined in the finite 
region, this finite region looks like the local solenoid
if one sees at infinity.
That is, $B |z_i|^2/4 \rightarrow \phi_c \ln|z_i| /2 \pi$ as $|z|
\rightarrow \infty$, where $\phi_c$ is the total magnetic flux
confined in this finite region induced by the uniform magnetic field.
Because of the symmetrizer ${\cal S}$, without loss of generality we 
can get one term in the symmetric form as the prototype for requiring 
the nomalizability of $\Psi_s$. We choose the term in an increasing
order $n^{'}_1 \le \cdots \le  n^{'}_N$.
Requiring that $\Psi_s$ be normalizable as one variable, say $z_i$,
gives the condition 
\begin{equation}
n^{'}_N + \alpha_s < \alpha_c - \theta (N-1) - 
\sum_{j (\ne i)} k_{ij} -1 ,
\end{equation}
where  $\alpha_c$ is defined by $\phi_c / \phi^*_0$.  
As two variables $z_i$ and $z_j$ approach the spatial infinity,
the condition of the normalizability becomes
\begin{equation}
n^{'}_{N-1} + n^{'}_N + 2 \alpha_s <
2 \alpha_c - \theta (N-1) - \theta (N-2) - \sum_{j (\ne i)} k_{ij}
- \sum_{l (\ne i,j)} k_{jl} -2.
\end{equation}
On the other hand, applying the increasing order to
Eq. (9) yields
\begin{equation}
n^{'}_{N-1} +n^{'}_N + 2 \alpha \le
2( n^{'}_N + \alpha_s) <
2 \alpha_c -2\theta (N-1) -2 \sum_{j (\ne i)} k_{ij} -2.
\end{equation}
The boundary condition (3) implies $\theta + k_{ij} >0$.
Comparing this with Eq. (11),  
it is easy to draw a conclusion that for $\theta + k_{ij}>0$ Eq. (9) 
alone is sufficient.
Combining this with $ n^{'}_N + \alpha_s>0$ from
the boundary condition gives the upper bound to the total
number $N$ of anyons. For this $N$ to be the greatest integer,
$0 < \theta + k_{ij} <1 $ for all $i$ and $j$. 
Since $\theta$ is the fixed statistical parameter, all the $k_{ij}$'s
must have the same value to satisfy $0 < \theta + k_{ij} <1 $.
Let us denote this $k_{ij}$ as $k_m$, then the total nomber $N$ is in 
the range
\begin{equation}
0<N< \frac{\alpha_c - (\alpha_s + n^{'}_N)
+ \theta +k_m -1}{\theta +k_m}.
\end{equation}
This equation implies that the zero energy ground state is
possible only for $\alpha_c>1$.
Then the maximum number of anyons in the ground state is determined by
\begin{equation}
N_{max} = \bigg[ \frac{\alpha_c - (\alpha_s + n^{'}_{N_{max}})
+ \theta +k_m -1}{\theta +k_m} \bigg],
\end{equation}
where $[\alpha]$ means the largest integer less than $\alpha$.
From the nature of multivalued wavefunctions (2), the statistics
is invariant under the parameter change $\theta \rightarrow
\theta + integer$. This periodic behavior agrees with the well-known
periodicity of the AB effect. And the statistics is also
invariant under the non-zero parameter change $\theta \rightarrow
- \theta$ for $\theta = integer$.
It is easily seen that $N_{max}$ is also invariant under the above
parameter changes, since we can change the integer $k_m$ so that
$\theta+k_m$ does not change.
In the region $0<\theta<1$, $N_{max}$ becomes the greatest number for
$k_m=0$.
With this $N_{max}$ anyons the corresponding wave function for 
$\alpha_s=0$ becomes
\begin{equation}
\prod_{i < j} (z_i - z_j)^{ \theta}
 \exp\left[- \frac{2 \pi B}{4 \phi^*_0} \sum_i |z_i|^2\right].
\end{equation}
This wave function represents the anyons completely filling 
the first Landau level.
For very large external flux $\alpha_c$, the maximum density in the unit 
of flux density is $N_{max} / (\phi_c /\phi_0)
\approx e^*/ (\theta e)$.
This agrees with the exact result commented in Ref. \cite{Kive2}.
When the external flux $\phi_c$ increases by one flux quantum $\phi_0$,
$\alpha_c$ will increase by $\alpha_c \rightarrow \alpha_c + 1/q$.
This implies one more quasiparticle to be available in the ground
state when $\phi_c$ is increased by $\phi_0$ for $\theta= 1/q$.
For general $\theta=p/q$, approximately $1/p$ anyons can be added to
the ground state for the increasement of the external flux by
one flux quantum.

For the type II anyons, only change in the Hamiltonian is 
$e^* \rightarrow e$.
Hence following the procedure similar to the type I anyons,
the ground-state wavefunction becomes
\begin{equation}
{\tilde \Psi}_s (z_1, \cdots, z_N)= \prod_i z_i^{{\tilde \alpha}_s}
\prod_{i < j} (z_i - z_j)^{ \theta} {\cal S} \prod_i z_i^{{\tilde n}_i}
\prod_{i<j} (z_i -z_j)^{ {\tilde k}_{ij}}
\exp\left[- \frac{2 \pi B}{4 \phi_0} \sum_i |z_i|^2 \right],
\end{equation}
where ${\tilde n}_i$ and ${\tilde k}_{ij}$ are integers respectively.
This ensures that the original wavefunction ${\tilde \Psi}_0$ is single-valued.
And ${\tilde \alpha}_s$ is defined as $\phi_s /\phi_0$.
The AB periodicity displays also the fundamental period.
From the normalizability of this wavefunction, the maximum number is
found to be
\begin{equation}
{\tilde N}_{max} = \left[ \frac{ {\tilde \alpha}_c - ({\tilde \alpha}_s
+ {\tilde n}_{N_{max}})+ \theta + {\tilde k}_m-1}{\theta +{\tilde k}_m} \right],
\end{equation}
where ${\tilde \alpha}_c$ is defined as $\phi_c / \phi_0$.
When $\phi_c$ is increased by $1$, the number of anyons
may increase approximately by $1 / \theta$ for the type II anyon.
For $0< \theta<1$ the corresponding wavefunction of the 
${\tilde N}_{max}$ anyons becomes the anyons completely filling the
first Landau level,
\begin{equation}
\prod_{i < j} (z_i - z_j)^{ \theta}
\exp\left[- \frac{2 \pi B}{4 \phi_0} \sum_i |z_i|^2\right].
\end{equation}
compare to Eq. (14), $\phi_0^*$ is replaced by $\phi_0$.

Next we consider that the two kinds of anyons of the same type 
with different statistical parameters $\theta^{'} (= p^{'} / q^{'})$ 
and $\theta(=p/q)$ are together in the presence of the external 
magnetic field.  
We consider the geometry, in which the two dimensional space contains 
an island of area $S$. We assume that the $\theta$ anyons reside only 
inside the island and that the $\theta^{'}$ anyons are outside only.
For simplicity, we also assume the area of the island is large enough
so that the above argument for the ground state remained valid.
And there is no additional impenetrable flux inside the island
except the one by the anyon itself.
The external uniform magnetic flux through the island is $\phi_c(=SB)$.
The ground state wavefunction inside the island with the maximum
number of anyons will be (14) and (17) for the type I and type 
II anyons respectively.
Since these wavefunctions are the same as the case of completely 
filling the first LL, we can consider the situation 
represented by these  wavefunctions as the equibrium situation.
The geometry is similar to what Jain {\it et. al.} \cite{Jain}
have considered, where a channel of $\nu^{'} = p^{'} /q^{'}$ FQHE 
liquid contains an island of the $\nu = p/q$ FQHE liquid.
We will consider the periodicty of the equilibrium with respect to
the external flux through the island.
When the magnetic flux inside the island is increased by one flux
quantum, $\phi_c \rightarrow \phi_c + \phi_0$, $1/p$ more anyons
with $\theta$ is required to restore the equilibrium for the case of 
type I.  Since we assume that there exist only two kinds of anyons in 
the system, the charge must be supplied by anyons with 
$\theta^{'}$ outside the island.
For the type I case we must have
$$
j \frac{1}{p} \frac{e}{q} = j^{'} \frac{e}{q^{'}} 
$$
to recover the equilibrium.
It is easily found the smallest $j$ is $pq/s$, where $s$ is the
greatest common factor of $pq$ and $q^{'}$.
It implies that the equilibrium will be periodic with $(pq/s) \phi_0$ 
period.  We note the periodicity found in Jain {\it et. al.} can
be obtained based on the above argument. 
Jain {\it et. al.} considered the situation 
when the channel liquid is $\nu_n$ FQHE liquid and the island liquid
is $\nu_{n-1}$ FQHE liquid, when $\nu_n (\equiv n/ (2n+1))$ is the 
filling factor of the principle FQHE liquid.
The liquid inside the island is considered in such a way that the 
quasiholes of the channel ($\nu_n$) FQHE liquid completely fill the 
lowest LL of their own.
That is, the wavefunction of the quasiholes in Ref. \cite{Jain} 
is the same as (14).
In our consideration this situation may be interpreted as follows.
The anyons with $\theta_n$ are in the equibrium inside the island.
And there are also the anyons with the same statistical parameter
outside the island, since the charge is supplied by the quasiholes
of $\nu_n$ FQHE liquid.  
In this case $q$ and $q^{'}$ are the same and equal to be $2n+1$.
And $p$ becomes $2n-1$.
Hence this system displays periodicity with the period $(2n-1) \phi_0$. 
It coincide with the result of Jain {\it et. al.} deduced from
the quasiclassical quantization of the anyon.

For the type II anyons, we need $q/p$ more anyons with $\theta$ in the
island to recover the equilibrium.
In this case the charges are the same for different statistical angles.
Hence there is no constraint of the charge conservation. 
Then the periodicity of the system is solely determined by the change of 
the maximum available number of anyons inside the island.
The system will recover its equilibrium when one additional anyon is 
available.  
If $\phi_c \rightarrow \phi_c + (p/q) \phi_0$, then
${\tilde N}_{max}$ becomes ${\tilde N}_{max} +1$. Therefore the 
periodicity of the system becomes $(p/q) \phi_0$. Then the system of 
type II anyons displays a fractional periodicity.
 
In conclusion, we have considered the exact ground state of two 
different types of anyons in the presence of external magnetic fields.
In case of the local impenetrable external flux, which
gives the true AB geometry, AB period is found to be the fundamental 
period in agreement with the general gauge invariance argument.
The maximum number of anyons in the ground state could be determined 
by the normalizability of the wavefunction.
When the total magnetic flux increases by one flux quantum $hc/e$,
one more anyon is available to the ground state for the type I anyon.
On the other hand, for the type II anyon the maximum number will
increase by $1/ \theta$.
We have also considered the geometry similar to that of Jain 
{\it et. al.}, in which a two-dimensional plane of $\theta^{'} (=p/q)$
anyons contains the island of $\theta (=p/q)$ anyons.
The equilibrium inside the island is shown to be periodic with respect
to the flux through the island. 
For the type I anyon the period equals the integer multiple of the 
fundamental flux, $(pq/s) \phi_0$, where $s$ is the greatest common 
factor of $pq$ and $q^{'}$.
The period for the type II anyon equals the fractional multiple,
$(p/q) \phi_0$.
We also could reproduce the result of Jain {\it et. al.}, by considering
the normalizability of the exact ground state wavefunction.

This work was supported in part by Korean Research Foundation, 
by POSTECH BSRI special fund and KOSEF.

\end{document}